\documentclass[%
 aip,
 amsmath,amssymb,
 reprint,
 twocolumn,
 dblfloatfix]{revtex4-2}
\usepackage{graphicx}
\usepackage{dcolumn}
\usepackage{bm}
\usepackage{siunitx}
\usepackage{amsmath}

\usepackage[usenames,dvipsnames,svgnames,table]{xcolor}
\colorlet{m}{black}
\colorlet{k}{Black}

\usepackage[textwidth=12.5mm]{todonotes}
\usepackage{easyReview}
\usepackage[colorlinks=true,citecolor=blue]{hyperref}
\usepackage[rightcaption]{sidecap}
\usepackage{cancel}

\begin{document}

\preprint{AIP/123-QED}

\title{Shubnikov-de Haas oscillations of two-dimensional electron gases in AlYN/GaN and AlScN/GaN heterostructures
}


\author{Yu-Hsin Chen}
\email{yc794@cornell.edu}
\affiliation{\hbox{Department of Materials Science and Engineering, Cornell University, Ithaca, NY, 14853, USA}}

\author{Thai-Son Nguyen}
\email{tn354@cornell.edu}
\affiliation{\hbox{Department of Materials Science and Engineering, Cornell University, Ithaca, NY, 14853, USA}}

\author{Isabel Streicher}
\affiliation{\hbox{Fraunhofer Institute for Applied Solid State Physics IAF, Tullastrasse 72, 79108 Freiburg, Germany}}
\affiliation{\hbox{Kavli Institute at Cornell for Nanoscale Science, Cornell University, Ithaca, NY, 14853, USA}}

\author{Jimy Encomendero}
\affiliation{\hbox{School of Electrical and Computer Engineering, Cornell University, Ithaca, NY, 14853, USA}}

\author{Stefano Leone}
\affiliation{\hbox{Fraunhofer Institute for Applied Solid State Physics IAF, Tullastrasse 72, 79108 Freiburg, Germany}}

\author{Huili Grace Xing}
\affiliation{\hbox{Department of Materials Science and Engineering, Cornell University, Ithaca, NY, 14853, USA}}
\affiliation{\hbox{Kavli Institute at Cornell for Nanoscale Science, Cornell University, Ithaca, NY, 14853, USA}}
\affiliation{\hbox{School of Electrical and Computer Engineering, Cornell University, Ithaca, NY, 14853, USA}}

\author{Debdeep Jena}
\email{djena@cornell.edu}
\affiliation{\hbox{Department of Materials Science and Engineering, Cornell University, Ithaca, NY, 14853, USA}}
\affiliation{\hbox{Kavli Institute at Cornell for Nanoscale Science, Cornell University, Ithaca, NY, 14853, USA}}
\affiliation{\hbox{School of Electrical and Computer Engineering, Cornell University, Ithaca, NY, 14853, USA}}

\begin{abstract}
AlYN and AlScN have recently emerged as promising nitride materials that can be integrated with GaN to form two-dimensional electron gases (2DEGs) at heterojunctions. Electron transport properties in these heterostructures have been enhanced through careful design and optimization of epitaxial growth conditions. In this work, we report for the first time Shubnikov-de Haas (SdH) oscillations of 2DEGs in AlYN/GaN and AlScN/GaN heterostructures, grown by metal-organic chemical vapor deposition. SdH oscillations provide direct access to key 2DEG parameters at the Fermi level: (1) carrier density, (2) electron effective mass ($m^* \approx 0.24\,m_{\rm e}$ for AlYN/GaN and $m^* \approx 0.25\,m_{\rm e}$ for AlScN/GaN), and (3) quantum scattering time ($\tau_{\rm q} \approx 68 \, \text{fs}$ for AlYN/GaN and $\tau_{\rm q} \approx 70 \, \text{fs}$ for AlScN/GaN). These measurements of fundamental transport properties provide critical insights for advancing emerging nitride semiconductors for future high-frequency and power electronics.

\end{abstract}

\maketitle

The formation of two-dimensional electron gases (2DEGs) in polar nitride heterostructures has enabled the development of high-electron mobility transistors (HEMTs)\cite{khan1992observation,asif1993high}, which are now widely used as commercial high-frequency RF amplifiers\cite{mishra2008gan,mishra2002algan} and fast high-voltage switching applications\cite{amano20182018,flack2016gan}. These devices are typically based on Al(Ga)N/GaN heterostructures.

More recently, 2DEGs have been realized in AlYN/GaN and AlScN/GaN heterostructures. Compared with conventional Al(Ga)N barriers, AlYN and AlScN barriers provide stronger spontaneous polarization\cite{Ambacher.2021b,Afshar.2025}, resulting in higher electron densities, and their larger conduction band offset further enhances 2DEG confinement.\cite{wang_band_2023, jin_band_2020} In addition, AlYN and AlScN can be lattice-matched to GaN, minimizing strain-induced degradation. Both AlYN and AlScN can be ferroelectric, making them attractive for non-volatile memory\cite{Fichtner.2019,Wang.2023d} and reconfigurable RF devices.\cite{Casamento2022_ferrohemts,Wang2023ferroelectricHEMT,Yang2023PulsedEDmode} These properties make AlScN/GaN and AlYN/GaN 2DEGs promising for next-generation high-speed, high-power GaN RF transistors.\cite{Green_2019_ScAlN-GaN_HEMT_2.4Amm, Tahhan_2022_PassivationSchemes_ScAlNbarrier,nomoto2025alscn,strainbalanceIEDM} Notably, metal–organic chemical vapor deposition (MOCVD) grown AlScN/GaN HEMTs have achieved a record output power of 8.4 W/mm with 42.0\% power added efficiency under class-AB continuous wave operation at 30 GHz (Ka-band).\cite{Krause.2023} Likewise, MOCVD-grown AlYN/GaN HEMTs have achieved a near-Boltzmann limit sub-threshold swing of 66.5 mV/dec.\cite{Nomoto.2025}

A high-quality 2DEG is evidenced by the occurrence of Shubnikov–de Haas (SdH) oscillations in longitudinal magnetoresistance \(R_{\rm xx}\). The condition for SdH oscillation is that the cyclotron energy \(\hbar \omega_{\rm c}\) exceeds both the thermal energy \(k_{\rm b}T\) and the Landau level broadening \(\hbar / \tau_{\rm q}\),\cite{sladek1958magnetoresistance} where \(\omega_{\rm c} = eB \, / \, m^*\) is the cyclotron frequency, \(k_{\rm b}\) is the Boltzmann constant, \(\tau_{\rm q}\) is the quantum scattering time, and \(m^*\) is the electron effective mass. To put it simply, \( B \times \mu_q\ > 1 \) is required, where \(\mu_q\) is quantum mobility. Consequently, the observation of SdH oscillations typically requires high magnetic fields, cryogenic temperatures, and high-quality 2DEGs with minimal scattering. SdH oscillations have been reported in GaN 2DEGs with AlGaN\cite{wong1998magnetotransport,elhamri19980,frayssinet2000high}, AlN\cite{cao2008very,chen2025shubnikov}, AlInN\cite{wang2018scatterings,karmakar2024enhanced} and AlInGaN\cite{karmakar2023enhanced} barriers. In contrast, SdH oscillations have not yet been reported for AlScN/GaN and AlYN/GaN 2DEGs. 


Formerly, the electron transport properties of MOCVD-grown AlScN/GaN and AlYN/GaN 2DEGs were limited by the unintentional graded AlScGaN or AlYGaN layer at the barrier/GaN channel interface, caused by the low growth rate and high growth temperature. This issue was mitigated by inserting a nominal AlN interlayer to  limit Al and Sc back-diffusion into the GaN channel, providing a sharper interface.\cite{Streicher.2023,Streicher.2024b} Further improvement in AlScN/GaN interface abruptness was achieved by using new Sc precursors with higher vapor pressures, which enabled faster AlScN growth rates. The resulting reduction in growth temperature suppressed atomic diffusion across the GaN interface and minimized impurity incorporation by shortening the growth time.\cite{Streicher.2024} Transport properties are also strongly impacted by alloy composition, with the high electron mobilities achieved for near-lattice-matched AlScN barriers containing $<$ 12\% Sc.\cite{Yassine.2025} These growth innovations have ultimately yielded room-temperature (RT) electron mobilities exceeding \(1100 \, \text{cm}^2/\text{Vs}\) in both AlScN/GaN\cite{streicher2023enhanced} and AlYN/GaN\cite{Streicher.2024b} 2DEGs with electron densities greater than \(1.2 \times 10^{13} \, \text{cm}^{-2}\), making them suitable for quantum transport studies.

In this work, we report for the first time well-resolved SdH oscillations in \(R_{\rm xx}\) for both AlYN/GaN and AlScN/GaN 2DEGs, observed at cryogenic temperatures and in magnetic fields up to 14 T. Analysis of these quantum oscillations yields (1) the carrier density, (2) the electron effective mass, and (3) the quantum scattering time, providing a direct evaluation of the transport properties in AlYN/GaN and AlScN/GaN 2DEGs.

\begin{figure}[ht]
	\centering
	\includegraphics[width=9cm]{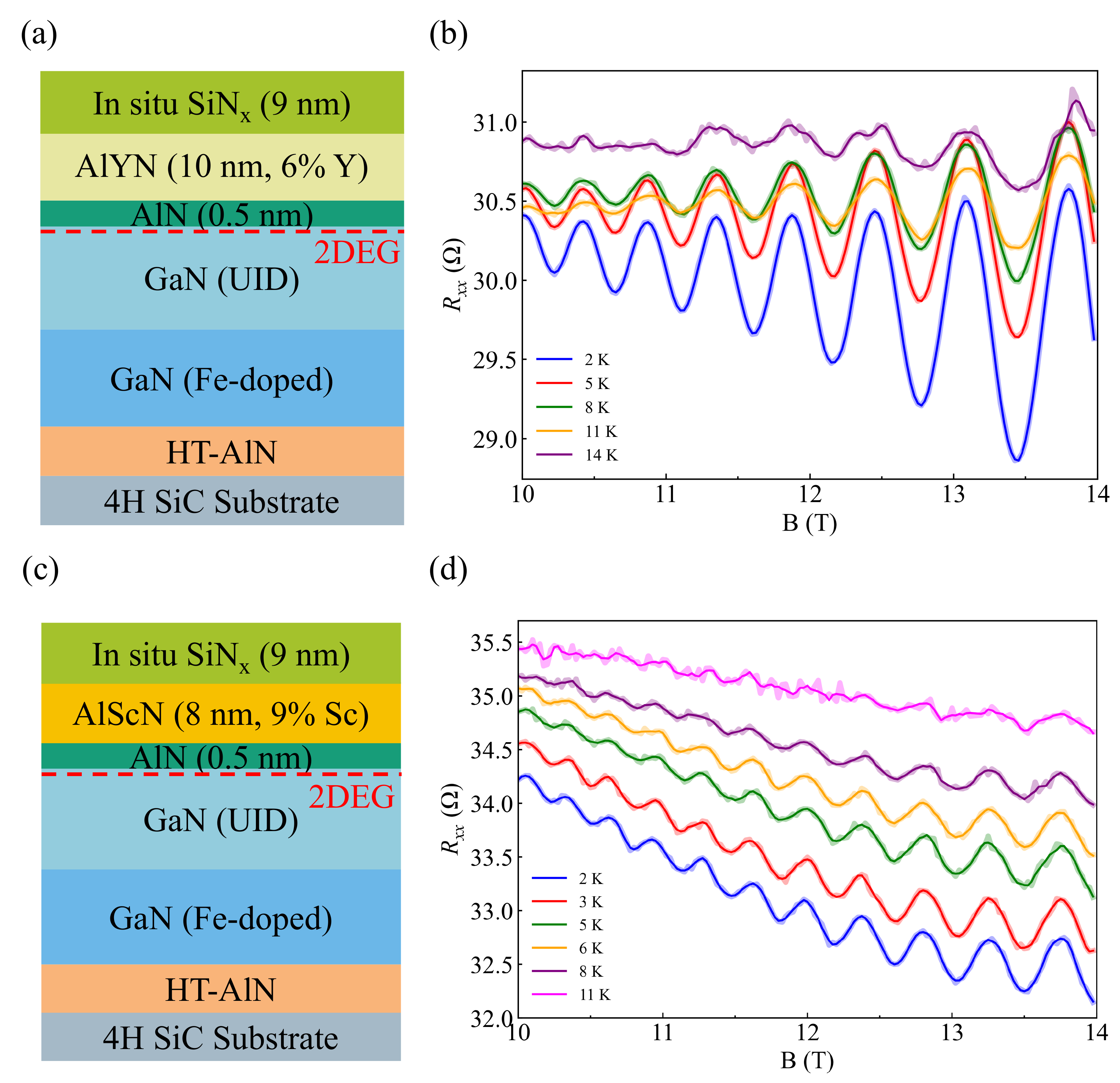} 
	\caption{\label{fig:1} Schematic diagrams of the epitaxial (a) AlYN/GaN and (c) AlScN/GaN heterostructures grown by MOCVD. Longitudinal magnetoresistance \(R_{\rm xx}\) of (b) the AlYN/GaN 2DEG and (d) the AlScN/GaN 2DEG, measured over temperatures from 2 to 11 K and magnetic field \(B\) from 10 to 14 T. The faint lines show the raw \(R_{\rm xx}\) data, while the solid lines represent the data obtained using a Savitzky–Golay filter, which reduces noise while preserving the peak positions and amplitudes.
	}
\end{figure}

Figures~\ref{fig:1}(a) and ~\ref{fig:1}(c) show the schematic cross-sections of the AlYN/GaN and AlScN/GaN heterostructures used in this work. Both AlScN/GaN and AlYN/GaN samples were grown on 4H SiC substrates by MOCVD. Fe-doped semi-insulating GaN buffer was grown on a high-temperature AlN nucleation layer on SiC, followed by an unintentionally doped (UID) GaN buffer and GaN channel. The barrier consists of a thin 0.5 nm AlN interlayer, followed by an AlYN (10 nm, 6\% Y) or AlScN (8 nm, 9\% Sc) barrier and an in-situ SiNx passivation layer of 9 nm to prevent oxidation of the barrier layers.\cite{manz_improved_2021,Wang.2023b,Leone.2023,Streicher.2025} The Sc and Y compositions were chosen for a near-lattice matched condition to GaN. The AlN interlayer is included in both structures to effectively boost the 2DEG mobility by providing a sharper interface, as previously reported.\cite{Streicher.2023,casamento_transport_2022} SIMS measurements reveal the presence of oxygen and carbon in the AlYN and AlScN barriers. The calculated energy band diagrams and electron density profiles for both AlYN/GaN and AlScN/GaN samples with these impurities are provided in Fig. S1 in the supplementary material. More details on the MOCVD growth conditions, the structural and transport characteristics of these samples can be found in previous publications.\cite{Streicher.2023,Streicher.2024,Streicher.2024b}

Low magnetic field Hall-effect measurements of the $1\ \text{cm}^{2}$ samples used for SdH measurements show that the AlYN/GaN sample hosts a 2DEG density of \(1.19 \times 10^{13} \, \text{cm}^{-2}\) with a mobility of \(1318 \, \text{cm}^2/\text{Vs}\) at RT, and increasing to \(3730 \, \text{cm}^2/\text{Vs}\) at 10 K. The $1\ \text{cm}^{2}$ AlScN/GaN sample for SdH measurements shows a 2DEG density of \(1.77 \times 10^{13} \, \text{cm}^{-2}\) with a mobility of \(1050 \, \text{cm}^2/\text{Vs}\) at 300 K. This AlScN/GaN sample used for SdH measurements accidentally broke into small pieces after the SdH measurements were performed. Therefore, low-temperature Hall measurements could not be performed on the same sample. Instead, the 10~K Hall measurement was carried out on another $1~\text{cm}^2$ sample taken from the same wafer, which yields a 2DEG density of $1.89 \times 10^{13}~\text{cm}^{-2}$ and a mobility of $1083~\text{cm}^2/\text{Vs}$ at 300~K, increasing to $2032~\text{cm}^2/\text{Vs}$ at 10~K.
 To more accurately represent the electron transport characteristics of the 100~mm wafers from which these $1\ \text{cm}^{2}$ samples were taken, the statistical transport mapping results measured by contactless Leighton instruments at 34 points across the wafer are presented in Table. S1 in the supplementary material.

\begin{figure*}
	\centering
	\includegraphics[width=0.992\textwidth]{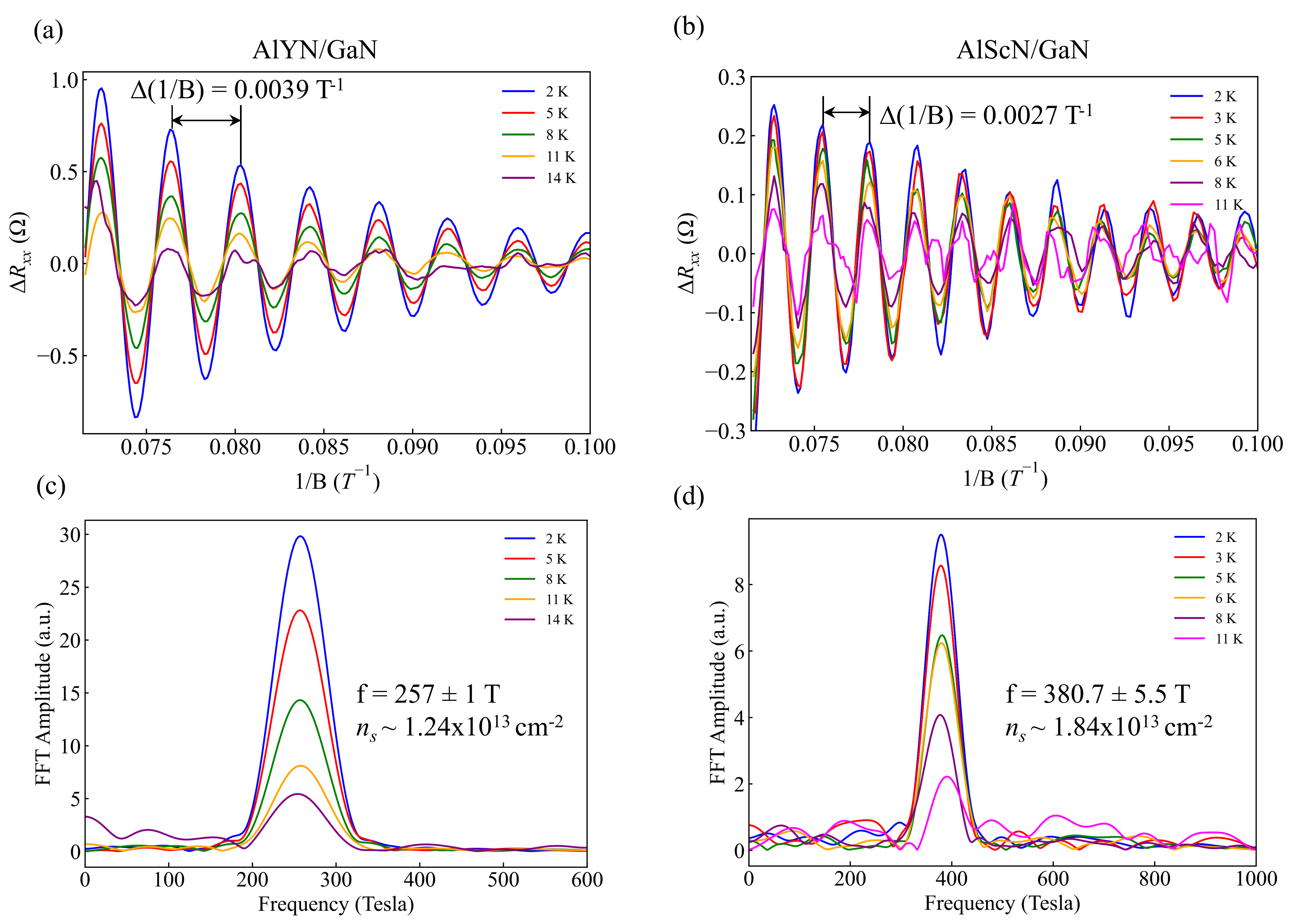} 
	\caption{\label{fig:2} The oscillatory component \(\Delta R_{\rm xx}\) as a function of \(1/B\) for (a) AlYN/GaN and (b) AlScN/GaN heterostructures.  For AlYN/GaN, the oscillation period \(\Delta(1/B) = 0.0039 \, \text{T}^{-1}\) corresponds to a 2DEG density of \(n_s \approx 1.23 \times 10^{13} \, \mathrm{cm}^{-2}\). In contrast, the smaller oscillation period of  \(\Delta(1/B) = 0.0027 \, \mathrm{T}^{-1}\)  in AlScN/GaN reflects a higher 2DEG density of \(n_s \approx 1.81 \times 10^{13} \, \mathrm{cm}^{-2}\).
    Fast Fourier transform (FFT) analysis of \(\Delta R_{\rm xx}\) vs. \(1/B\) shows a single dominant frequency for both samples. The 2DEG densities extracted from both real-space oscillation period and reciprocal-space FFT analyses are consistent with low-field Hall effect measurements.}
\end{figure*}

\begin{figure*}
	\centering
	\includegraphics[width=0.7\textwidth]{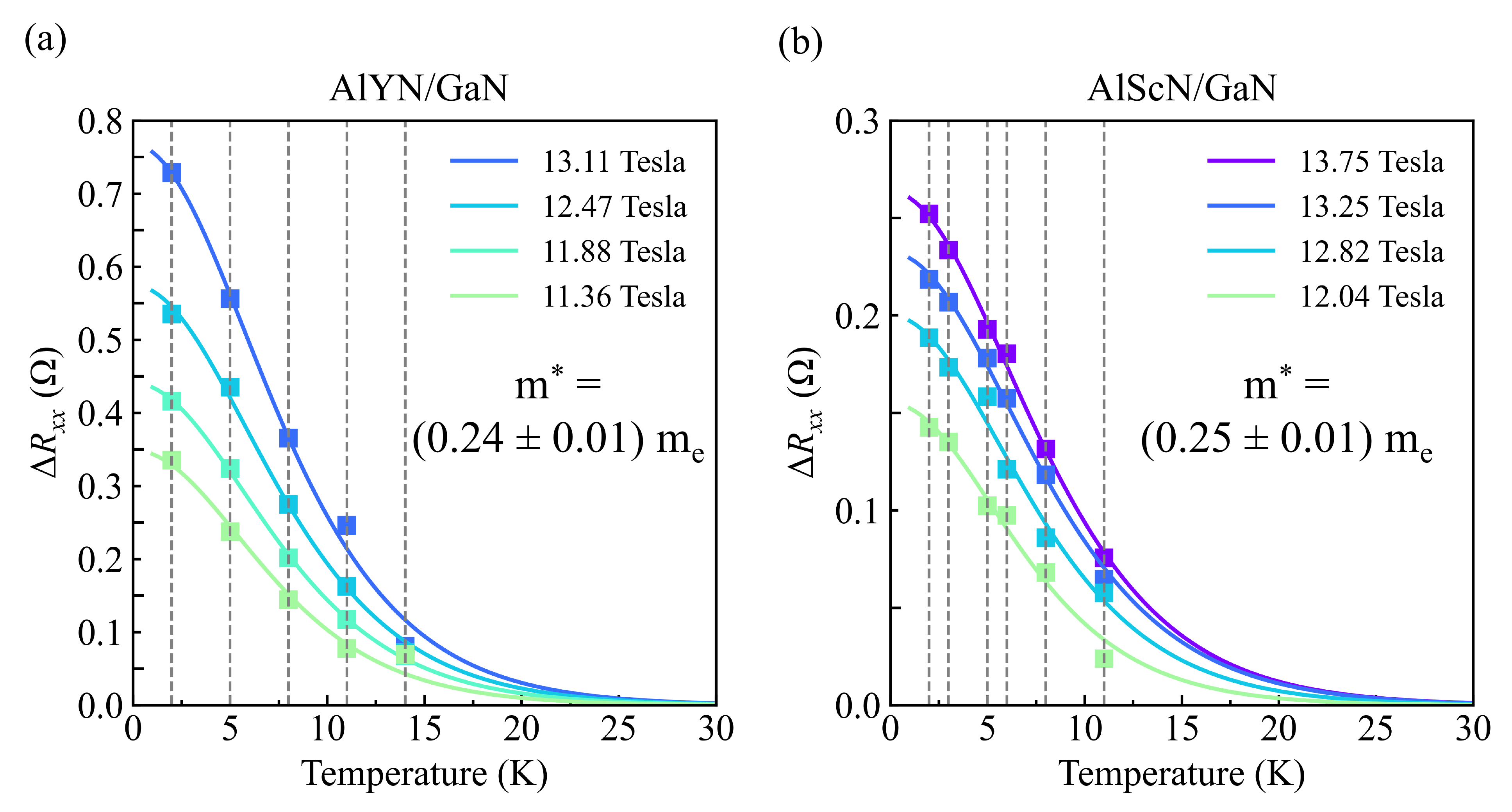}
	\caption{\label{fig:3} Temperature dependence of peak \(\Delta R_{\rm xx}\) at various B fields for (a) AlYN/GaN and (b) AlScN/GaN samples. The solid curves represent the fits to \(\chi / \sinh(\chi)\) in Eq.(1) to extract the electron effective mass, yielding \( m^* = (0.24 \pm 0.01) \, m_{\rm e} \) for the AlYN/GaN 2DEG and \( m^* = (0.25 \pm 0.01) \, m_{\rm e} \) for the AlScN/GaN 2DEG.}
\end{figure*}


Magnetotransport measurements were carried out on both heterostructures using 1 cm $\times$ 1 cm coupons, with indium-soldered ohmic contacts formed in a Van der Pauw configuration. The samples were measured in a Physical Property Measurement System (PPMS) from Quantum Design under a DC excitation current of \(100 \, \mu\text{A}\) and magnetic field \(B\) up to 14 T. Figures.~\ref{fig:1}(b) and~\ref{fig:1}(d) show the measured longitudinal magnetoresistance \(R_{\rm xx}\) as a function of the magnetic field \(B\) for the AlYN/GaN 2DEG and AlScN/GaN 2DEG, respectively. The faint lines show the raw \(R_{\rm xx}\) data, while the solid lines represent the data obtained using a Savitzky–Golay filter, which reduces noise while preserving the peak positions and amplitudes. All subsequent analyses were performed on the filtered data. Magnetic fields between 10 and 14 T were applied perpendicular to the sample surface, and the measurements were repeated at temperatures ranging from 2 to 14 K. The \(R_{\rm xx}\) oscillation amplitude increases with magnetic field as the Landau level separation increases and each level accommodates more states. In contrast, the oscillation amplitude is dampened at higher temperatures due to increased Fermi–Dirac distribution smearing near the Fermi level. The
onsets of the SdH oscillations were recorded at around 8-9 T for both
samples, with the most pronounced oscillations observed at higher magnetic fields and lower temperatures.

The oscillatory component \(\Delta R_{\rm xx}\) was obtained by subtracting the background resistance from the measured \(R_{\rm xx}\). For both samples, the background was determined by averaging two separate fourth-order polynomial fits—one to the peaks and another to the valleys of the SdH oscillations\cite{okazaki2018shubnikov}, as shown in Fig. S2. \(\Delta R_{\rm xx}\) is expressed as  \cite{kubo1959quantum,dingle1952some,ihn2009semiconductor,jena2022quantum} 
\begin{equation}
\Delta R_{\rm xx} \propto \frac{\chi}{\sinh(\chi)}e^{-\frac{\pi}{\omega_{\rm c} \tau_{\rm q}}}\cos\left(\frac{2\pi E_{\rm F}}{\hbar \omega_{\rm c}}\right),
\end{equation} where \(\chi = 2\pi^2 k_B T / (\hbar \omega_{\rm c})\) and \(E_{\rm F} = n_{\rm s} \pi \hbar^2 / m^*\) is the Fermi level of the 2DEG. \(\Delta R_{\rm xx}\) is periodic in inverse magnetic field \(1/B\), which is determined by the cosine term in Eq. (1).  \(\Delta R_{\rm xx}\) provides a direct measurement of the 2DEG density via the Onsager relation \( n_{\rm s} = (2 q / h) \times 1 / \Delta(1/B) \). Figures~\ref{fig:2}(a) and \ref{fig:2}(b) show \(\Delta R_{\rm xx}\) versus \(1/B\) for both the AlYN/GaN and AlScN/GaN 2DEGs. The oscillation period is extracted as \(\Delta(1/B) = 0.0039 \, \mathrm{T}^{-1}\) for the AlYN/GaN 2DEG, corresponding to a 2DEG density of \(n_s \approx 1.23 \times 10^{13} \, \mathrm{cm}^{-2}\). For the AlScN/GaN 2DEG, the smaller oscillation period, \(\Delta(1/B) = 0.0027 \, \mathrm{T}^{-1}\), corresponds to a higher 2DEG density of \(n_s \approx 1.81 \times 10^{13} \, \mathrm{cm}^{-2}\).

Figures~\ref{fig:2}(c) and~\ref{fig:2}(d) present the fast Fourier transform (FFT) analysis of $\Delta R_{\rm xx}$ versus $1/B$ for both samples. Prominent frequency peaks of $f = 257 \pm 1~\mathrm{T}$ and $f = 380.7 \pm 5.5~\mathrm{T}$ were obtained for the AlYN/GaN and AlScN/GaN 2DEGs, respectively. The FFT frequency can be used to calculate the 2DEG density via the relation, $n_{\rm s} = \left( 2q / h \right) \times f$, yielding $n_{\rm s} \approx 1.24 \times 10^{13}~\mathrm{cm}^{-2}$ for AlYN/GaN and $n_{\rm s} \approx 1.84 \times 10^{13}~\mathrm{cm}^{-2}$ for AlScN/GaN. The 2DEG densities extracted from both real-space $\Delta(1/B)$ and reciprocal-space FFT analyses are consistent with the low-field Hall-effect measurements. 

The effective mass of the 2DEG at the Fermi level can be determined from temperature-dependent SdH oscillations measured at a fixed B field. Figures ~\ref{fig:3}(a)\ and ~\ref{fig:3}(b) shows the damping of the peak \(\Delta R_{\rm xx}\) with increasing cryogenic temperature at various B fields for both samples. For each B field, the data were fitted to the thermal damping factor \(\chi/\sinh(\chi)\) in Eq.(1), yielding an electron effective mass of \( m^* = (0.24 \pm 0.01) \, m_{\rm e} \) for AlYN/GaN 2DEG and \( m^* = (0.25 \pm 0.01) \, m_{\rm e} \) for AlScN/GaN 2DEG, where \( m_{\rm e} \) is the free electron mass. The increased $m^{*}$ compared to bulk GaN is likely due to a combination of band nonparabolicity\cite{syed2003nonparabolicity} and electron--phonon (polaron) coupling \cite{wu1997cyclotron}. Since the electron wavefunction in AlScN/GaN and AlYN/GaN heterostructures is primarily confined within the GaN channel with slight penetration into the AlN interlayer [See Fig. S2], the polaron effect is expected to be similar to that in AlGaN/GaN. Given that the 2DEG density at AlScN/GaN and AlYN/GaN interfaces typically exceeds $n_s$=\(10^{13} \, \text{cm}^{-2}\), electron--phonon screening is strong, suggesting that band nonparabolicity is likely the dominant contributor to the enhanced electron effective masses.

The values extracted from SdH oscillations in this work are in good agreement with recently published data obtained using terahertz optical Hall effect measurements at 40 K, which show \( m^* = 0.2-0.23 \, m_{\rm e} \) for AlYN/GaN with $n_s \sim 1 \times 10^{13}~\text{cm}^{-2}$ and \( m^* = 0.25-0.27 \, m_{\rm e} \) for AlScN/GaN with $n_s$=\(2-2.7 \times 10^{13} \, \text{cm}^{-2}\).\cite{Stanishev.2025} Fig. S3(a) in the supplementary material shows a comparison of experimentally reported values of $m^{*}$ extracted from SdH oscillations as a function of $n_s$ in AlYN and AlScN barriers of this work with those reported for AlGaN, AlN, and AlIn(Ga)N barrier heterostructures in prior reports.

\begin{figure}[ht]
	\centering
	\includegraphics[width=0.8\columnwidth]{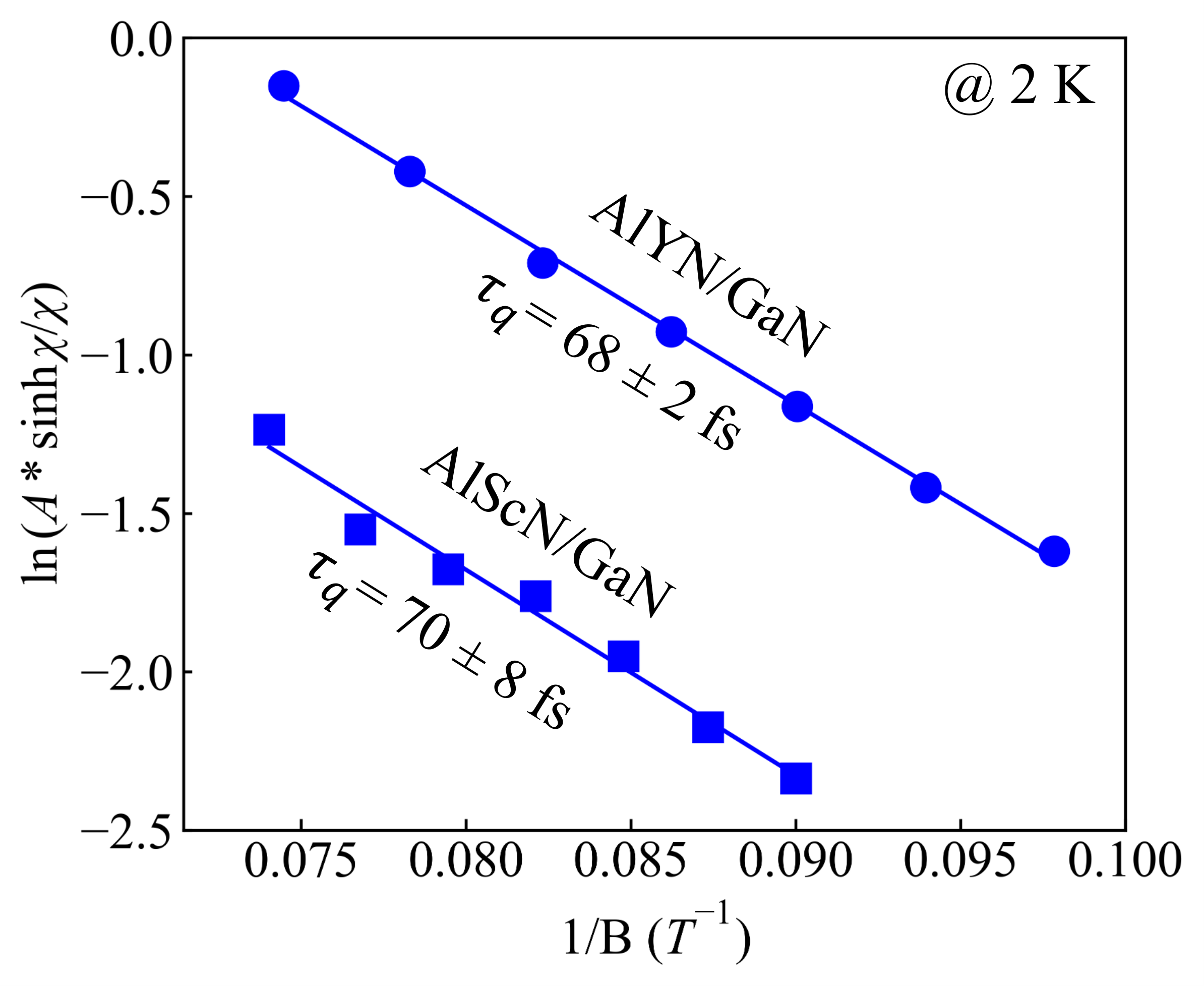}
	\caption{Dingle plots, where \(\ln\left(A^* \sinh(\chi) \, / \, \chi\right)\) is plotted versus \(1/B\) at 2 K, for both AlYN/GaN and AlScN/GaN 2DEG, allowing for the extraction of quantum scattering time \( \tau_{\rm q} \). \( A^* \) denotes the absolute value at the valley of \(\Delta R_{\rm xx}\). The solid black lines are fits to the disorder damping term \(e^{-\pi / (\omega_{\rm c} \tau_{\rm q})}\) from Eq.(1).}
	\label{fig:4}
\end{figure}

Landau levels exhibit a finite broadening due to their coupling with scattering potentials arising from structural disorder. This broadening is quantified by the quantum scattering lifetime \( \tau_{\rm q} \), which governs the exponential decay of the \( \Delta R_{\rm xx} \) amplitude with increasing 1/B at a fixed temperature. For a given temperature, \( \tau_{\rm q} \) is extracted by fitting the disorder-related damping term \(e^{-\pi / (\omega_{\rm c} \tau_{\rm q})}\) from Eq.(1) to the Dingle plot, where \(\ln\left(A^* \sinh(\chi) \, / \, \chi\right)\) is plotted versus \(1/B\) and \( A^* \) denotes the absolute value at the valley of \(\Delta R_{\rm xx}\). The slope from the linear fit of a Dingle plot is equal to $-\pi m^* / (q \tau_{\rm q})$, from which one can calculate \( \tau_{\rm q} \). For both samples, the low-frequency components were filtered out, leaving only the main FFT peak. An inverse FFT was then applied to this FFT peak to reconstruct the oscillations, which were subsequently used for the Dingle analysis. As shown in Fig.~\ref{fig:4}, \( \tau_{\rm q} \) is determined to be $\tau_{\rm q} = 68\pm2\,\text{fs}$ for the AlYN/GaN 2DEG and $\tau_{\rm q} = 70\pm8\,\text{fs}$ for AlScN/GaN 2DEG. 

While the quantum scattering lifetime \( \tau_{\rm q} \) has no dependence on the scattering angle, the momentum scattering time \( \tau_{\rm m} \) is weighted by the factor of \( 1/(1 - \cos \theta) \), with \( \theta \) denoting the scattering angle.\cite{harrang1985quantum} For isotropic scattering the ratio $\tau_m/\tau_q$ goes to unity, and values much larger than unity indicate anisotropic scattering mechanisms. The momentum scattering time \( \tau_{\rm m} \) is extracted from the low-field Hall mobility using the Drude relation \(\mu = e \tau_{\rm m} / m^*.\) At 10 K, Hall-effect measurement yields \( \tau_{\rm m}\) = 496 fs for the AlYN/GaN 2DEG and the Dingle ratio is \( \tau_{\rm m}/\tau_{\rm q} \sim 7.3 \). For AlScN/GaN, the 10~K Hall-effect result obtained from the the second $1~\text{cm}^2$ sample yields $\tau_{m} = 289~\text{fs}$ and a Dingle ratio of $\tau_{m}/\tau_{q} \sim 4.1$. The Dingle ratio $> 1$ indicates the presence of long-range scattering potentials.

The Dingle value indicates the presence of long-range scattering potentials. A potential candidate is Coulomb scattering\cite{jena2002quantum}, possibly arising from charged dislocations\cite{jena2000dislocation}, charge centers like impurities in the barriers\cite{frayssinet2000high, elhamri19980}, or microscopic strain-induced defects from the barrier layers, whose long range strain fields reach the channels.\cite{sumiya2025strain} Due to a 3.4\% lattice mismatch, heteroepitaxial growth of wurtzite GaN on hexagonal 4H-SiC results in a high threading dislocation density of 10$^8$--10$^{10}$ cm$^{-2}$. In addition, high oxygen and carbon levels have been detected in the MOCVD-grown AlScN and AlYN barrier layers\cite{streicher2023enhanced,Leone.2023}, and the influence is reflected in the measured capacitance–voltage characteristics.\cite{Yassine.2025} Jena et al. reported that among Coulombic scattering mechanisms, the Dingle ratio is substantially larger for dislocation scattering than for impurity scattering.\cite{jena2002quantum} Fig.~S3(b) in the supplementary material compares the $\tau_q$ reported here with 2DEGs in barriers other than AlYN and AlScN. Reduction of Coulomb scattering (e.g. by dislocations and barrier impurities) can improve the quantum scattering time in future AlYN/GaN and AlScN/GaN structures.

In summary, we report the observation of SdH oscillations in GaN 2DEGs with novel AlYN and AlScN barriers, grown by MOCVD. The oscillation periodicity provides a direct measurement of the 2DEG density, in agreement with low-field Hall-effect measurements. Analysis of the thermal damping yields the electron effective mass, with \( m^* = (0.24 \pm 0.01) \, m_{\rm e} \) for the AlYN/GaN 2DEG and \( m^* = (0.25 \pm 0.01) \, m_{\rm e} \) for the AlScN/GaN 2DEG. Furthermore, analysis of the Dingle plot reveals quantum scattering time of $\tau_{\rm q} = 68\pm2\,\text{fs}$ for the AlYN/GaN 2DEG and $\tau_{\rm q} = 70\pm3\,\text{fs}$ for the AlScN/GaN 2DEG. These results demonstrate quantum transport in AlScN/GaN and AlYN/GaN heterostructures and lay the foundation for future investigations into the materials science, fundamental physics, and device applications of these emerging nitride semiconductors.


\begin{acknowledgments}
This work is supported by Army Research Office under Grant No. W911NF2220177; and SUPREME, one of seven centers in JUMP 2.0, a Semiconductor Research Corporation (SRC) program sponsored by DARPA. This work made use of the Cornell Center for Materials Research shared instrumentation facility which are supported through the NSF MRSEC program (DMR-1719875), and Kavli Institute at Cornell (KIC). 

The development of MOCVD growth was funded by the BMBF Project ProMat\_KMU "PuSH" Grant Number 03XP0387B. 
The Fraunhofer IAF authors thank Dr. Lutz Kirste and Dr. Patrik Stra\v{n}\'{a}k for the structural characterization of the MOCVD-grown heterostructures, which were published previously\cite{Streicher.2024,Streicher.2024b}.

\end{acknowledgments}

\section*{Supplementary Material}
See the supplementary material for (1) the calculated energy band diagrams
and electron density profiles for both AlYN/GaN and
AlScN/GaN samples, (2) the background resistance of SdH oscillations, and (3) comparisons of experimentally reported electron effective mass $m^{*}$ and quantum scattering time \( \tau_{\rm q} \) extracted from SdH oscillations in AlYN and AlScN barriers of this work with those reported for AlGaN, AlN, and AlIn(Ga)N barrier heterostructures in prior reports.

\section*{Author Declarations}
\subsection*{Conflict of Interest}
The authors have no conflicts to disclose.

\section*{Data Availability}
The data that support the findings of this study are available from the corresponding author upon reasonable request.

\section*{References}
\bibliography{main}

\end{document}